\documentstyle[preprint,amsfonts,aps]{revtex}

\begin{document}
\draft
\preprint{HEP/123-qed}
\title{Rigorous Proof of a Liquid-Vapor Phase Transition\\
in a Continuum Particle System
}
\author{J.L. Lebowitz\cite{grant}}
\address{Department of Mathematics and Physics, Rutgers University,
New Brunswick, NJ 08903, USA\\
lebowitz@math.rutgers.edu}
\author{A.E. Mazel\cite{grant}}
\address{Department of Mathematics, Rutgers University,
New Brunswick, NJ 08903, USA\\
International Institute of Earthquake
prediction Theory and Theoretical Geophysics, 113556 Moscow, Russia\\
mazel@math.rutgers.edu}
\author{E. Presutti}
\address{Dipartimento di Matematica, 
Universit\`a di Roma Tor Vergata, Via della Ricerca Scientifica, 
00133 Roma, Italy\\
presutti@axp.mat.utovrm.it}
\maketitle
\begin{abstract}
We consider particles in ${\Bbb R}^d, d \geq 2$, interacting
via  attractive pair and  repulsive four-body potentials  of the Kac
type.  Perturbing about mean field theory, valid when the interaction range
becomes infinite, we prove rigorously the existence of a liquid-gas phase
transition when the interaction range is finite but long compared to the
interparticle spacing.
\end{abstract}


An outstanding problem in equilibrium statistical mechanics is to derive
rigorously the existence of liquid-vapor phase transition in particles
interacting with any kind of reasonable potential, say Lennard-Jones or
hard core plus attractive square well. 
This is in marked contrast to the
situation for lattice systems where proofs of phase transitions abound.
Thus for an Ising model with ferromagnetic interactions in dimensions $d
\geq 2$ there are known to be two coexisting phases at low temperatures.
These are perturbations of the two ground states, namely the configuration
with all spins up or all spins down.  At non-zero temperature, there are
fluctuations which cause the formation of droplets of the opposite phase,
but their energy cost is so high that they remain, at low temperatures, in
$d \geq 2$, only small perturbations of the ground state.  It was Peierls
\cite{[1]} who first gave a convincing argument of the validity of such a
picture; the argument was later made fully rigorous by Dobrushin
\cite{[2]} and Griffiths \cite{[3]}.  Independently of this general
argument, Onsager \cite{[4]} solved the two dimensional Ising model on
${\Bbb Z}^2$, with nearest neighbor interactions, explicitly and found the
behavior of the system near the critical temperature which marks the end
point of phase coexistence. Since that time solutions have been found for
many other two dimensional lattice models \cite{[5]}. At the same time
ferromagnetic and other inequalities as well as the development of the
powerful Pirogov-Sinai formalism \cite{[6]} have resulted in a
comprehensive rigorous theory of phase transitions in lattice systems, in
$d \geq 2$, at sufficiently low temperatures.

The extension of these results to continuum particle systems has proven
difficult.  The ground states of such systems are not at all easy to
characterize; they are presumed to be periodic or quasi-periodic
configurations which depend in some complicated way on the interparticles
forces.  This is however far from proven and hence the analysis of the
fluctuations that appear when we increase the temperature above zero are
correspondingly harder, indeed very much harder, to study than in the
simple lattice systems; moreover key inequalities like the ferromagnetic
ones are no longer available.  These problems have been overcome so far
only for some multicomponent systems with special features.  In
particular, Ruelle, \cite{[7]}, proved that the two component
Widom-Rowlinson model \cite{[8]} has a demixing phase transition.
Ruelle's proof strongly exploits the symmetry between the components
present in this model: see also later proofs of phase transitions in
related models \cite{[9]}, \cite{[10]}, \cite{[11]}.  There are also
proofs of phase transitions in $d=1$ for continuum systems with
interactions which decay very slowly or not at all.  Such models with many
particle interactions were analyzed by Fisher and Felderhof \cite{[12]},
while Johanson \cite{[13]} has considered pair interactions which decay as
$r^{-\alpha}$, $1 < \alpha < 2$, proving that at low temperatures the
pressure is not differentiable.

In this letter we report the first 
proof of a liquid-vapor transition in one-component continuum systems with
finite range interactions and no symmetries.  The basic idea of our
approach is to study perturbations not of the ground state but of the mean
field behavior (mfb), i.e.\ we shall consider situations where the
interactions are parametrized by their range $\gamma^{-1}$ \cite{[14],[15]}
and perturb about $\gamma = 0$ which gives mfb.  Such scaling potentials
were investigated by Kac, Uhlenbeck and Hemmer [KUH] \cite{[16]} for a
system of one dimensional hard rods with an added pair potential
\begin{equation}
\phi_{\gamma}(q_i,q_j)=- \alpha {1 \over 2} \gamma \exp
[-\gamma |q_i -q_j|],\quad \gamma, \alpha >0.
\label{1.1}
\end{equation}
This was later generalized by Lebowitz and Penrose [LP] \cite{[17]} to
$d$-dimensional systems with suitable short range interactions and general
Kac potentials of the form
\begin{equation}
\phi_{\gamma}(q_i,q_j)= -\alpha \gamma^d J(\gamma |q_i - q_j|)
\label{1.2}
\end{equation}
with $\int_{{\Bbb R}^d} J(r) dr =1, \quad J(r) >0$

LP showed that in the infinite volume limit followed by the limit $\gamma
\to 0$ the Helmholtz free energy $a$ takes the form,
\begin{equation}
\lim_{\gamma \to 0} a(\rho, \gamma)=C E\{ a_0(\rho)-
 {1 \over 2} \alpha \rho^2 \}
\label{1.4}
\end{equation}
Here $\rho$ is the particle density, $a_0$ is the free energy density of
the reference system, i.e. the system with $\alpha=0$ in
(\ref{1.2}). $a_0$ is convex in $\rho$ (by general theorems) and $CE\{
f(x)\}$ is the largest convex lower bound of $f$.  (The dependence of
$a_0$ on the temperature $\beta^{-1}$ has been suppressed.) For $\alpha$
large enough the term in the curly brackets in (\ref{1.4}) has a double well
shape and the $CE$ corresponds to the Gibbs double tangent
construction. This is equivalent to Maxwell's equal area rule applied to a
van der Waals' type equation of state where it gives the coexistence of
liquid and vapor phases \cite{[17]}.

Following the work of KUH and LP, various attempts were made to go
beyond the $\gamma = 0$ limit \cite{[18]}, \cite{[19]}.  It is clear
from general arguments, and it follows also explicitly from
\cite{[14]}, that in $d=1$ there is no phase transition for $\gamma >
0$.  Straightforward expansions in $\gamma$ are therefore bound to
fail for $d=1$, in the two phase region.  In $d>1$ these schemes give
plausible, but uncontrolled, approximations.  The main difficulty
comes from the fact that the phase transition is a singular event,
whose dependence on the parameter $\gamma$ is not at all smooth.  To
overcome this problem requires all the modern machinery of
Pirogov-Sinai theory built up in the past twenty five years [6, 20,
21] plus considerable additional effort.

It is the success of such an effort which enables us to 
show for some systems in $d \geq 2$, that
their behavior at finite $\gamma >0$ are close to mfb at $\gamma =0$, so
that a phase transition in the latter yields a phase transition in the
former for sufficiently small $\gamma$.  Such results have been recently
obtained for Ising models, \cite{[20]}, where
one uses a version of the Peierls argument, exploiting the spin flip
symmetry of the model.  The absence of symmetries in our case requires
instead the whole machinery of the Pirogov-Sinai theory \cite{[6]}. To
insure stabilization against collapse, which would be induced by a Kac
attractive pair potential, the natural choice made by KUH and LP is to
replace point particles by hard spheres or similar strongly repulsive pair
interactions. Our approach however does not work in such a case, as we
need a cluster expansion for the unperturbed reference system (i.e.
without the Kac interaction) at values of the chemical potential or
density for which it is not proven to hold.  Instead we consider point
particles and insure stability by introducing a positive four body
potential of the same range as the attractive two body one.  The
unperturbed system is then the free, ideal gas for which the cluster
expansion holds trivially.  The price is a much more involved mean field
analysis, which requires a special choice of the form of the interactions.

We now specify the model, state precisely our results, and give a flavor
of the proof \cite{[22]}.  Let $q = \{q_i\}, i=1,2,...$ be a configuration
of particles in a domain $\Lambda \subset {\Bbb R}^d, d \geq 2$.  The
energy of the configuration $q$ is
\begin{eqnarray}
\label{1.10}
&&H_\gamma(q) = 
- {1\over 2!}\sum_{i_1} \sum_{i_2 \not= i_1}
J_\gamma^{(2)}(q_{i_1},q_{i_2})  \\
+&&{1\over 4!} \sum_{i_1} \sum_{i_2 \not= i_1}
\quad \sum_{i_3 \not= i_1,i_2}
\quad \sum_{i_4 \not= i_1,i_2,i_3} 
J_\gamma^{(4)}(q_{i_1},q_{i_2},q_{i_3},q_{i_4}). \nonumber
\end{eqnarray}
Here
\begin{equation}
J_\gamma^{(2)}(q_{i_1},q_{i_2})=\gamma^d J^{(2)}
(\gamma q_{i_1},\gamma q_{i_2}),
\end{equation} 
\begin{equation}
J_\gamma^{(4)}(q_{i_1},q_{i_2},q_{i_3},q_{i_4})=
\gamma^{3d} J^{(4)}(\gamma q_{i_1},\cdots,\gamma q_{i_4}),
\end{equation} 
and
\begin{equation}
J^{(2)}(r_1,r_2) \geq 0$, $J^{(4)}(r_1,r_2,r_3,r_4) \geq 0,
\end{equation}
are fixed, bounded, translation invariant functions of finite range: they
vanish whenever any of the distances $|r_i -r_j|$ is larger than some  fixed
length $l_d$.

The equilibrium properties of this system are specified by a grand
canonical ensemble with reciprocal temperature $\beta$, chemical potential
$\lambda$, and suitable boundary conditions (bc), i.e.\ by the Gibbs
measure $\mu_{\Lambda, \gamma, \beta, \lambda} ^{bc}$.  To prove
coexistence of liquid and vapor phases for some $\beta$ and $\lambda$ we
have to show that by choosing two different bc, one favoring the liquid
and another the vapor phase, call them $+$ and $-$, the Gibbs measures
obtained in the limit $\Lambda \nearrow {\Bbb R}^d$, describe two phases
differing primarily by their densities; the appropriate order parameter
for this transition.  We do this in detail \cite{[21]} for a particular
choice of the interactions
\begin{eqnarray}
&&J^{(2)}(r_{1},r_{2})=  |B (r_{1}) \cap B(r_{2})|,\\ 
&&J^{(4)}(r_{1},r_{2},r_{3},r_{4})=  |\cap_{j=1}^4 B (r_{j})|
\label{1.7}
\end{eqnarray}
where $B(r)$ is the ball in ${\Bbb R}^d$ of volume 1 and center $r$, i.e.\
$J^{(2)}(r_1, r_2)$ is equal to the overlap volume of the two balls (of
radius $\pi^{-1/2}$ and $(4/3 \pi)^{-1/3}$ in $d = 2, 3$) centered at
$r_1$ and $r_2$.  Similarly $J^{(4)}(r_1, r_2, r_3, r_4)$ is equal to the
overlap volume of four such balls.

\medskip
\noindent{\bf Theorem.} 
\nobreak
{\sl Let $\beta_c = ({3 \over 2})^{{3 \over 2}}$ and $\beta_0 > \beta_c$
(as below), then, for any $\beta \in (\beta_c,\beta_0)$ there exist
functions $\gamma_0(\beta)$ and $\lambda(\gamma, \beta)$ such that for $0
<\gamma < \gamma_0(\beta)$ the model with $J^{(2)}$ and $J^{(4)}$ as in
(\ref{1.7}) has at least two distinct infinite volume Gibbs measures
$\mu^{\pm}_{\gamma,\beta}$.  These measures are translation invariant and
ergodic (with respect to space translations), with an exponential decay of
correlations. They have particle densities respectively equal to
$\rho_{\gamma,\beta,-}>0$ and $\rho_{\gamma,\beta,{+}} >
\rho_{\gamma,\beta,{-}}$.  In the limit $\gamma \to 0$,
$\lambda(\gamma,\beta) \to \lambda(\beta)$, $\rho_{\gamma,\beta,\pm} \to
\rho_{\beta,\pm}$ and there exist positive constants $c$ and $\delta$ such
that, $|\lambda(\gamma,\beta) - \lambda(\beta)| + \sum_{s=\pm}
|\rho_{\gamma,\beta,s} - \rho_{\beta,s}| \le c \gamma^\delta$.  }

\medskip
The reason for the particular choice of the interactions (\ref{1.7}) as
well as for the appearance of $\beta_0$ in the Theorem are related to the
mfb of the system (\ref{1.10}) valid when $\gamma \to 0$ (following the
limit $\Lambda \nearrow {\Bbb R}^d$).  The mean field equilibrium profiles
are functions $\rho^\star(r)$ that minimize the mean field Gibbs free
energy functional,
\begin{eqnarray}
\label{1.5}
{\cal F}_{\beta,\lambda}&&(\rho)= 
\int  dr  {\rho(r)\over \beta}(\log\rho(r) -1) \\
- &&\int  dr \lambda\rho(r) 
-{1\over 2!} \int dr_1 dr_2 J^{(2)}(r_1,r_2) \rho(r_1) \rho(r_2) \nonumber
\\ 
+&&{1\over 4!} \int  dr_1 \cdots dr_4 J^{(4)}(r_1, \dots, r_4)
\rho(r_1) \cdots \rho(r_4) \nonumber
\end{eqnarray}
where $\rho(r)$ is a test density profile and the first integral gives the
entropy contribution to the free energy.

The minimizers $\rho^\star(r)$ satisfy the mean field equation $\delta
{\cal F}_{\beta,\lambda}/$ $\delta \rho(r) = 0$.  The resulting equation
is highly non linear and not very much is known about its solutions for
general $J^{(2)}$ and $J^{(4)}$.   For the particular choice (\ref{1.7})
Eq. (\ref{1.5}) simplifies to
\begin{eqnarray}
\label{1.5.1}
{\cal F}_{\beta,\lambda}(\rho) &&= \int dr[{\rho(r) \over \beta} (\log
\rho(r)-1) - \lambda R(r,\rho) \nonumber \\
&&- {1 \over 2!} R^2(r,\rho) + {1 \over 4} 
R^4(r,\rho)]
\end{eqnarray}
where $R(r,\rho)$ is the average density over a ball of unit volume in
${\Bbb R}^d$ centered at $r$.  It is now easy to show, using the convexity
of the first term, that 
 the minimizers are always spatially
homogeneous i.e., they corespond to a constant density $\rho \geq 0$.
For such a density,  the Gibbs free energy per
unit volume given in (\ref{1.5}) takes the form
\begin{equation}
f(\rho) = \beta^{-1}
\rho(\log \rho - 1) - \lambda \rho - \rho^2/2 + \rho^4/24.  
\label{1.6}
\end{equation}
The minimizing density will therefore be a solution of the equation
\begin{equation}
f^\prime(\rho) = \beta^{-1} \log \rho - \lambda - \rho + \rho^3/6 = 0.
\label{1.8}
\end{equation}
This equation has a unique solution for all $\lambda$ when $\beta \le
\beta_c = (3/2)^{3/2}$, while for $\beta > \beta_c$ there exists a
$\lambda(\beta)$ such that there are two minimizing solutions
$\rho_{\beta,+} > \rho_{\beta, -}$ with $\lambda(\beta)$ and
$\rho_{\beta,\pm}$ the same as in the last statement of the Theorem.  In
other words, $f(\rho)$ is convex for $\beta \le \beta_c$ and has a double
well shape of equal height for $\beta > \beta_c$, $\lambda =
\lambda(\beta)$. Moreover, there is a $\beta_0 > \beta_c$, given by the
smallest value of $\beta > \beta_c$ for which $f''(\rho_{\beta,\pm}) =
2(\beta \rho_{\beta,\pm})^{-1}$, such that the diagonal part of ${\delta^2
{\cal F}_{\beta,\lambda(\beta)} \over \delta \rho(r)\delta \rho(r')} \bigg
|_{\rho=\rho_{\beta,\pm}}$ is positive and dominates the non diagonal
ones.  When $\beta>\beta_0$ the second variational derivative of ${\cal
F}_{\beta,\lambda(\beta)}$ is still positive at $\rho = \rho_{\beta,\pm}$
but the diagonal part no longer dominates.  The former case is much
simpler to analyze and we have so far only worked out all the details for
that case.

To prove our Theorem we carry out a controlled Pirogov-Sinai cluster
expansion about the $\gamma = 0$ mean field state.  The first step in this
analysis is a ``coarse graining'', in which we partition space into cubes
of size $l$, with $l$ very large compared to the interparticle spacing but
small compared to $\gamma^{-1}$.  Given a particle configuration $q$ we
call $\{ \rho_x \}$, $x$ the centers of the cubes, the particle densities
in each cube.  We then show that the measure over the $\{\rho_x\}$,
obtained by integrating out all the other variables is, to within
controllable errors,  a Gibbs measure  with an effective Hamiltonian which is
essentially a discrete version of the mean field free energy functional
(\ref{1.5}) with $\rho(r)$ there replaced by $\rho_x$, and  balls
replaced by 
``lattice balls'' .  The important effect
of this procedure is that the new effective inverse temperature is $\beta
l^d$.  For $\gamma$ small enough and $l$ correspondingly large enough, we
are now in the right setup for the Pirogov-Sinai theory.   The remainder
terms are exponentially decaying multibody interactions. 

The ``ground states'' of our ``lattice system'' corresponding to the
vapor and liquid states are now defined by ensembles of configurations
having the $\rho_x$ ``close'' to the mean field vapor and liquid
densities $\rho_{\beta,+}$ and $\rho_{\beta,-}$.  The analysis of this
system is conceptually close to the one used in the extension of
Pirogov-Sinai theory to continuous (unbounded) spin systems developed
in \cite{[21]}.  Our analysis is actually simpler than that in
\cite{[21]}.  Instead of using a cluster expansion which requires
dealing with interactions among many Peierl's type contours separating
``bubbles'' of one ground state inside another, brought about by the
extended range of the potentials, we use a more analytic approach.  We
show in particular that the restricted effective Hamiltonian giving
the Gibbs measures of the lattice system corresponding to the $+$ or
$-$ ground state ensembles satisfy the Dobrushin uniqueness condition.
We then show that this remains true even after the addition of
contours to the ground states.  {}From this follows the exponential
decay of correlations in the liquid and vapor phases, for $\beta_c <
\beta < \beta_0$, stated in the Theorem.  We expect to prove that
similar results will hold even at lower temperatures, $\beta \geq
\beta_0$, but, as already mentioned, the proof is now more difficult:
technically, the equation satisfied by the stationary points of the
free energy functional (\ref{1.5}) is no longer a contraction and the
criteria for Dobrushin uniqueness is no longer satisfied by the
lattice system.


\begin{references}
\bibitem[*]{grant}   Work supported in part by NSF
grant DMR 95-23266 and AFOSR grant 4-26435
\bibitem{[1]} R.Peierls, 
{\it Proc. Camb. Phil. Soc.} {\bf 32}:477 (1936).
\bibitem{[2]} R.L.Dobrushin, 
{\it Th. Prob. Appl.} {\bf 10}:193 (1965).
\bibitem{[3]} R.B.Griffiths,
{\it Phys. Rev.} {\bf 136A}:437 (1964).
\bibitem{[4]} L.Onsager,
 {\it Phys. Rev.} {\bf 65}:117 (1944).
\bibitem{[5]} R.J.Baxter, {\it Exactly solved models in statistical
mechanics}, London-New York: Academic Press (1982).
\bibitem{[6]} S.A.Pirogov and Ya.G.Sinai, 
{\it Theor. and Math. Phys.}  {\bf 25}:358, {\bf 25}:1185 (1975).
\bibitem{[7]} D.Ruelle, 
{\it Phys. Rev. Let.}  {\bf 27}:1040 (1971).
\bibitem{[8]} B.Widom and
J.S.Rowlinson, {\it J. Chem. Phys.} {\bf 52}:1670 (1970).
\bibitem{[9]} J.L.Lebowitz and E.H.Lieb, 
{\it Phys. Let.} {\bf 39A}, N2:98 (1972).
\bibitem{[10]} J.Bricmont, K.Kuroda
and J.L.Lebowitz, {\it Comm. Math. Phys.} {\bf 101}:501 (1985).
\bibitem{[11]} H.-O. Georgii and O. Haggstrom, {\it Comm. Math. Phys.} {\bf
181}:507 (1996).
\bibitem{[12]} B.U.Felderhof
and M.E.Fisher, {\it Annals of Phys.}  {\bf 58} N1:176, N1:217, N1:268
(1970).
\bibitem{[13]} K.Johansson, 
{\it Comm. Math. Phys.}  {\bf 169}:521 (1995).
\bibitem{[14]} M. Kac, {\it Phys. Fluids} {\bf 2}:8 (1959).
\bibitem{[15]} G. A. Baker, Jr., {\it Phys. Rev.} {\bf 126}:2071 (1962).
\bibitem{[16]} M.Kac, G.Uhlenbeck and 
P.C.Hemmer, {\it J. Mat. Phys} {\bf 4}:216, {\bf 4}:229 (1963), {\it
J. Mat. Phys} {\bf 5}:60 (1964).
\bibitem{[17]} J.L.Lebowitz and O.Penrose,
{\it J. Mat. Phys} {\bf 7}:98 (1966).
\bibitem{[18]} M. Kac and C. J. Thompson, {\it Proc. Nat. Acad. Sci.}, {\bf
55}, 676 (1966) and {\it J. Math. Phys.}, {\bf 10}, 1373 (1969);
A. J. E. Siegert, C. J. Thompson and D. J. Vezzetti, {\it J. Math. Phys.},
{\bf 11}, 1018 (1970); E. Helfand, in {\it The Equilibrium Theory of Classical
Fluids}, III--41, H. L. Frisch and J. L. Lebowitz, eds., N.A. Benjamin,
1964.
\bibitem{[19]} J. L. Lebowitz, G. Stell, and S. Baer, {\it J. Math. Phys.},
{\bf 6}, 1282 (1965); G. Stell, J. L. Lebowitz, S. Baer, and W. Theumann,
{\it J. Math. Phys.}, {\bf 7}, 1532 (1966); see also section 4 of the
article by P. C. Hemmer and J. L. Lebowitz, in {\it Phase Transitions and
Critical Points}, C. Domb and M. S. Green, editors, {\bf 5B}, Academic
Press, London, 1976.
\bibitem{[20]} M.Cassandro and E. Presutti,
{\it Markov Processes and Related Fields} {\bf 2}: 241 (1996);
A. Bovier and M. Zahradnik, {\it J. Stat. Phys.} to appear; 
T.Bodineau and E.Presutti,
{\it Comm. Math. Phys.} {\bf 189}:287 (1997).
\bibitem{[21]} R.L.Dobrushin and M.Zahradnik,
in {\it Mathematical Problems of Statistical Mechanics and Dynamics},
R.L.Dobrushin ed., Dordrecht, Boston: Kluwer Academic Publishers, 1
(1986).
\bibitem{[22]} J.L.Lebowitz, A.E.Mazel and 
E.Presutti, preprint, (1997).

\end{references}
\end{document}